\documentclass[lettersize]{article}

\usepackage{PRIMEarxiv}

\usepackage[utf8]{inputenc} % allow utf-8 input
\usepackage[T1]{fontenc}    % use 8-bit T1 fonts
\usepackage{hyperref}       % hyperlinks
\usepackage{url}            % simple URL typesetting
\usepackage{booktabs}       % professional-quality tables
\usepackage{amsfonts}       % blackboard math symbols
\usepackage{nicefrac}       % compact symbols for 1/2, etc.
\usepackage{microtype}      % microtypography
\usepackage{fancyhdr}       % header
\usepackage{graphicx}       % graphics
\graphicspath{{media/}}     % organize your images and other figures under media/ folder
\usepackage{amsmath}
\usepackage{graphicx}
\usepackage{pdflscape}
\usepackage{rotating} % Rotating table
\usepackage{lmodern}
\usepackage{pdflscape}
\usepackage{lmodern}
\usepackage{array, multirow}

\usepackage{afterpage}
\usepackage{capt-of}% or use the larger `caption` package
\usepackage{multirow}
\usepackage{color}

\title{A NOVEL GPR-BASED PREDICTION MODEL FOR CYCLIC BACKBONE CURVES OF REINFORCED CONCRETE SHEAR WALLS}
\author{Zeynep Tuna Deger, Ph.D. and Gulsen Taskin Kaya, Ph.D. \\
  Istanbul Technical University \\
  \texttt{\{zeynep.tuna@itu.edu.tr, gulsen.taskin@itu.edu.tr\}} \\
}

\begin{document}
\maketitle

\begin{abstract}

 Backbone curves are used to characterize nonlinear responses of structural elements by simplifying the cyclic force – deformation relationships. Accurate modeling of cyclic behavior can be achieved with a reliable backbone curve model. In this paper, a novel machine learning-based model is proposed to predict the backbone curve of reinforced concrete shear (structural) walls based on key wall design properties. Reported experimental responses of a detailed test database consisting of 384 reinforced concrete shear walls under cyclic loading were utilized to predict seven critical points to define the backbone curves, namely: shear at cracking point ($V$\textsubscript{cr}); shear and displacement at yielding point ($V$\textsubscript{y} and $\delta$\textsubscript{y}); and peak shear force and corresponding displacement ($V$\textsubscript{max} and $\delta$\textsubscript{max}); and ultimate displacement and corresponding shear ($V$\textsubscript{u} and $\delta$\textsubscript{u}). The predictive models were developed based on the Gaussian Process Regression method (GPR), which adopts a non-parametric Bayesian approach. The ability of the proposed GPR-based model to make accurate and robust estimations for the backbone curves was validated based on unseen data using a hundred random sampling procedure. The prediction accuracies (i.e., ratio of predicted/actual values) are close to 1.0, whereas the coefficient of determination (R${}^{2}$) values range between 0.90 – 0.97 for all backbone points. The proposed GPR-based backbone models are shown to reflect cyclic behavior more accurately than the traditional methods, therefore, they would serve the earthquake engineering community for better evaluation of the seismic performance of existing buildings. 
\end{abstract}


\section{Introduction}
\subsection{Background on Shear Wall Modeling}

Reinforced concrete shear walls are commonly used to resist earthquake loading to provide substantial lateral strength and stiffness. As they are designed to undergo inelastic deformations under high-intensity ground shaking, it is essential to understand the nonlinear behavior and reflect the hysteretic responses in analytical models as accurately as possible. There are various well-accepted modeling approaches in the literature (e.g. summarized in NIST GCR 17-917-45 \cite{NISTATC}), that can be classified into two main approaches, namely: finite element and member modeling. The finite element modeling approach includes a detailed establishment of the problem by typically describing constitutive relations, and geometric nonlinearity is at the stress-strain level. Despite its advantages, such as being able to yield comprehensive results and capture local responses, the finite element modeling approach is typically cumbersome and computationally expensive for large-scale modeling and post-processing. Member modeling approaches, i.e., macro-models, are considered to be more convenient from this point of view, particularly for practitioner engineers, due to their ease-of-use and computational efficiency.

Numerous macro-models are available in the literature that can make reasonably accurate predictions for the structural response of reinforced concrete shear walls. Macro-models are also referred to as discrete finite element models in the literature \cite{taucer1991fiber}. Among macro-models, distributed and lumped plasticity models are those that have been adopted by various seismic design standards (e.g., ASCE 41-17 \cite{american2017asce}, Eurocode 8 \cite{standard2005eurocode}) and implemented into various analysis software \cite{Perform3D,etabs,mazzoni2006opensees} for modeling of reinforced concrete shear walls. The distributed plasticity model (also referred to as the fiber model) discretizes wall cross-section using nonlinear uniaxial fiber elements, assuming that plane sections remain plane after loading, whereas the lumped plasticity approach adopts the use of elastic elements with effective stiffness and zero-length plastic hinges at regions where material nonlinearity is expected to occur (that is, typically member ends). When compared to the fiber model, the lumped plasticity model has several disadvantages, such as neglecting (i) deformations at section neutral axis during loading-unloading, (ii) contribution of horizontal elements (e.g., beams, slabs) that are connected to the walls, (iii) influence of the change in axial load on wall shear strength and stiffness. Nevertheless, the method stands out with its ease-of-use and computational efficiency in practice. It should also be noted that both models have a common disadvantage, that is, shear-flexure interaction is not taken into account as coupled axial–flexural (i.e., P-M) behavior is represented by stress-strain relations of concrete and reinforcing steel along with generally elastic or elastoplastic horizontal spring to reflect the shear behavior. Other modeling techniques in literature account for shear-flexure interaction based on strut and tie method \cite{vulcano1988analytical,panagiotou2012nonlinear,fischinger2012modeling,kolozvari2019state} and multiple vertical line element model \cite{fischinger2012modeling,kolozvari2015modeling,kolozvari2019shear}; however, they are out of the scope of this paper.

\begin{figure}[!h]
\centering
\includegraphics[scale=0.4]{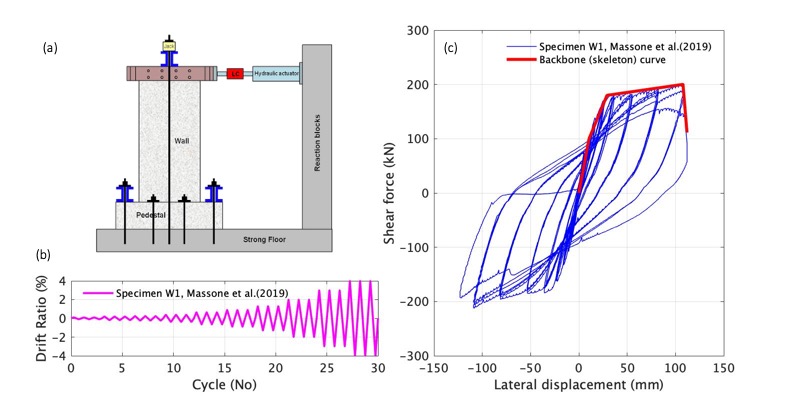}
\caption{(a) Reversed cyclic test setup; (b) loading protocol; (c) load – deformation response \cite{massone2019experimental}.}
\label{fig:1}
\end{figure}

Zero-length plastic hinges in the lumped plasticity model can be employed in two forms: (i) rotational springs, (ii) fiber sections. This paper focuses on the use of rotational springs. Nonlinear static and dynamic analyses are conducted based on such plastic hinge models characterized by force-deformation relationships, also known as "backbone curves", to ensure load and deformation paths follow but not exceed them. Backbone curves are constructed by connecting peak points at the first cycle of each increment \cite{american2017asce}, whereas they can be approximated by a series of linear segments broken at critical points for the sake of practicality (Fig. \ref{fig:1}).

\begin{figure}[!h]
\centering
\includegraphics[scale=0.6]{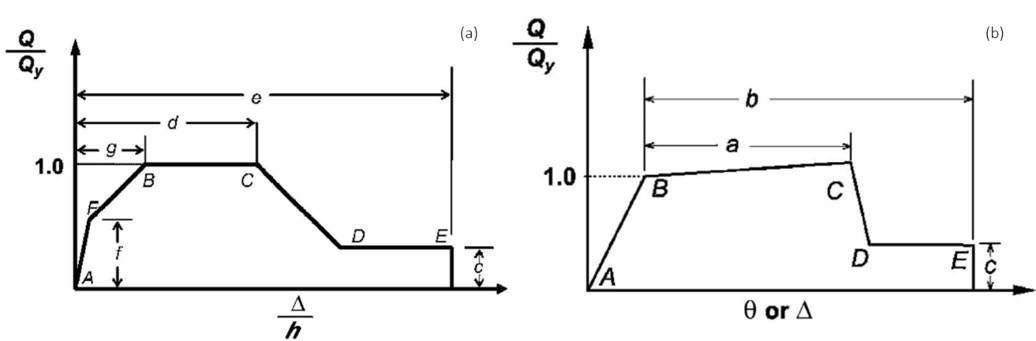}
\caption{Generalized force-deformation relations based on ASCE 41-17 for a) shear-controlled walls, b) flexure-controlled walls.}
\label{fig:ascebc}
\end{figure}

The history of the lumped plasticity approach dates back to the 1960s \cite{clough1966effect,clough1967nonlinear,giberson1967response,takeda1970reinforced}, whereas the earliest introduction of the use of backbone curves in a seismic performance-based evaluation guideline was in 1997 \cite{fema273,fema274}. In later years, cyclic backbone curves for shear walls were recommended in FEMA 356 (2000)    \cite{fema356} and ASCE 41 (2006) \cite{american2006asce}, which updated the modeling parameters based on the relevant research studies (Wallace et al. for squat walls \cite{wallace2006lightly}). Definition of backbone curves and associated modeling parameters for reinforced concrete shear walls in the most recent version of ASCE 41 \cite{american2017asce} are presented in Fig.\ref{fig:ascebc}. The multilinear backbone curve model recommended for shear-controlled walls includes five points at cracking (F), yielding (B), peak strength (C), ultimate displacement (D), and residual strength (E). The backbone curve model for flexure-controlled shear walls (Fig.\ref{fig:ascebc}(b)) is less refined (excludes the cracking point) as the deformations related to shear are negligible compared to the deformations associated with flexure, whereas the -x axis represents rotation (versus drift) for such walls. The normalized shear force ($Q$/$Q$\textsubscript{y}) and drift ratio ($\Delta$/h) values at such points are defined according to wall axial load level (as well as shear stress level for flexure-controlled walls), where normalized shear force at cracking is constant and equal to ($V$\textsubscript{cr}/ $V$\textsubscript{y} = 0.6) where $V$\textsubscript{y} equals to $V$\textsubscript{n}, i.e., the nominal shear strength calculated based on ACI 318 \cite{aci2019building}. It is worth noting that the definition of backbone curves in ASCE 41 was very similar for reinforced concrete columns up until ASCE 41-13 \cite{american2013asce}, where column modeling parameters were used to obtain based on a combination of column axial load level, reinforcement ratio, and shear stress ratio depending on the column failure mode. However, a major revision has been made in ASCE 41-17, which now provide parameters in the form of equations rather than the past table form. This revision suggests that predictive equations ensure better accuracy and easier use, whereas a similar study is needed for reinforced concrete shear walls, in parallel with this study.
Nonlinear modeling of reinforced concrete shear walls based on lumped plasticity approach has been investigated by various researchers for walls with different failure types. The lumped plasticity approach was addressed in terms of plastic hinge length \cite{bohl2011plastic,birely2012seismic,kazaz2012suneklik} and line-element models \cite{pugh2015nonlinear} for flexure-controlled walls, whereas semi-empirical equations were developed for backbone curves of low-rise shear-controlled walls by Carrillo and Alcocer \cite{carrillo2012backbone}. More recently, Epackachi et al. addressed the ASCE 41 cyclic backbone curve definitions by recommending equations for peak shear strength for shear-controlled walls \cite{epackachi2019cyclic,NISTATC}.

As summarized above, traditional  approaches are available in the literature to define backbone curves of reinforced concrete shear walls; however, they either require interpolation and generally provide conservative lower-bound estimates with unknown dispersion (ASCE 41-17) or do not provide the entire behavior characteristics (e.g., equations for lateral strength only \cite{NISTATC}). Not to mention, traditional  approaches do not provide recommendations on modeling parameters for walls with shear-flexure interaction (according to ASCE 41-17 \cite{american2017asce}, 1.5 $\le$ $h$\textsubscript{w}/$h$\textsubscript{w} $\le$ 3). This study aims to fill this gap by approaching the problem from a machine-learning (ML) point of view, which has not been explored before, and to propose predictive models to obtain cyclic backbone curves of RC shear walls based on key wall design properties (concrete compressive strength, horizontal boundary reinforcing ratio, etc). The prediction of the backbone curves is essentially a multi-output regression problem. Although the use of machine learning methods in the field of structural analysis and material modeling date back to the late 1980s \cite{adeli1989perceptron}, \cite{vanluchene1990neural}, such methods have received a lot of attention in the field of structural earthquake engineering \cite{salehi2018damage}, in particular, for modeling the physical relations of a structural system as an input-output problem via clustering, classification, or regression \cite{Sudret2014b}, \cite{Jiang2015a}. More recently, the machine learning methods have been utilized to predict earthquake-induced building damage \cite{Zhang2018}, to define backbone curve models for reinforced concrete columns \cite{luo2018machine}.

\section{Experimental database}
\begin{figure}[!h]
\includegraphics[scale=0.035]{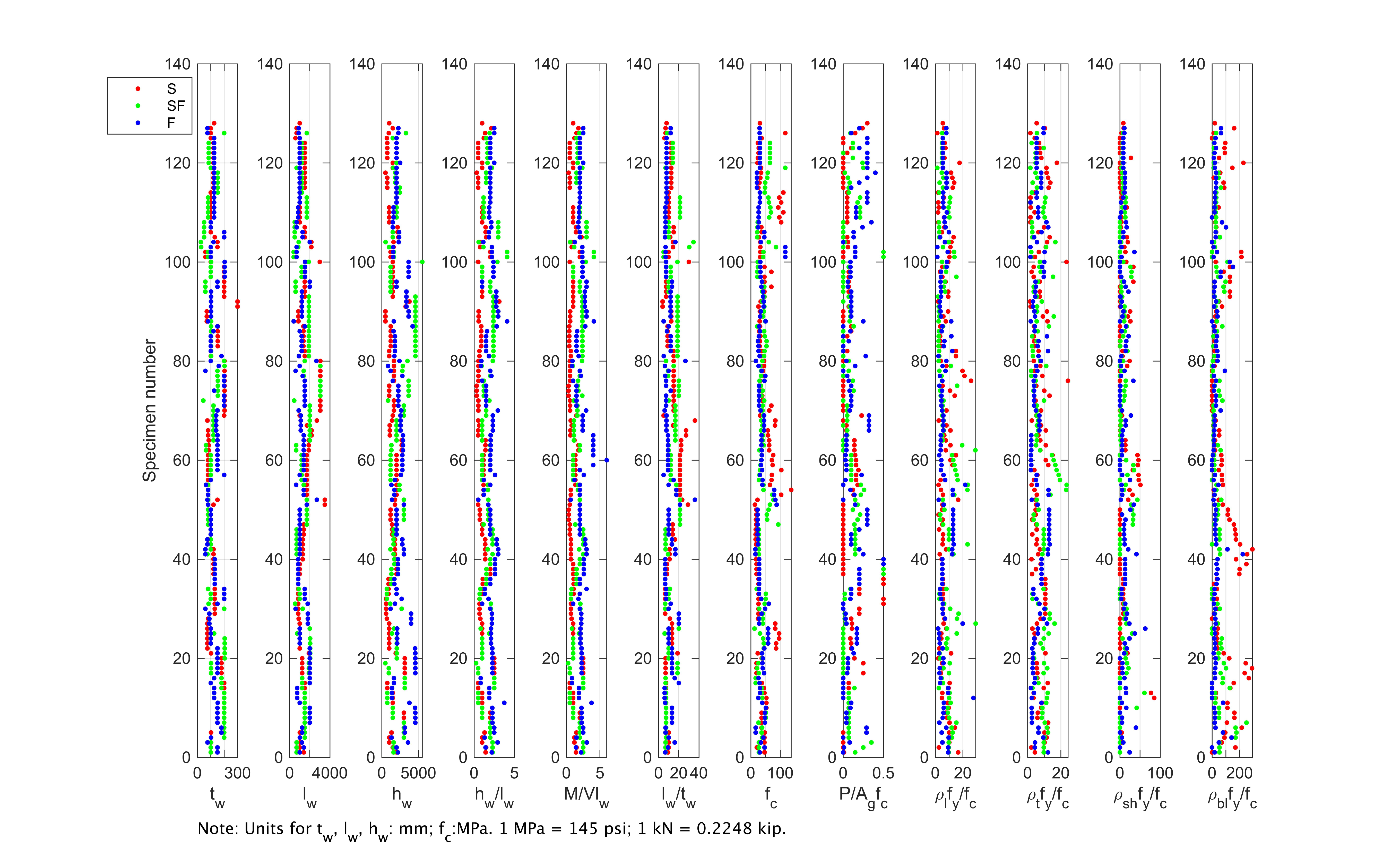}
\caption{Distribution of key wall properties (384 specimens)}
\label{fig:distributions}
\end{figure}

To develop predictive models for shear wall backbone curves based on the machine learning approach, the key wall properties (Fig.\ref{fig:distributions}) and the critical backbone variables are designated as input variables (features) and output variables, respectively. To achieve this, more than 500 shear wall specimens from various experimental studies conducted worldwide are collected based on a review of available research \cite{deger2019empirical,NEESHub,grammatikou2015strength}. Repaired or strengthened walls, specimens tested under monotonic loading, as well as specimens constructed with diagonal web reinforcement, confinement devices, composite materials (e.g., I-section confinement), or hollow sections are excluded, resulted in 384 specimens useful for this study. 

  The database includes specimens that have different cross-sections (rectangular [255/384] or barbell- shaped or flanged [129/384]), different end conditions (cantilever [341/384] or rotations fixed on top [43/384]), and different failure types (130 shear-controlled specimens, 126 specimens with shear-flexure interaction, and 128 flexure-controlled specimens). Shear-controlled specimens are those reported to fail in diagonal tension failure or web crushing by reaching their shear strength before flexural yielding occurs, whereas flexure-controlled specimens are those that yield in flexure before reaching their shear strength, thus show damages such as concrete spalling and crushing and/or reinforcing bar buckling at the boundary elements.  Specimens that contain both damage types are classified as walls controlled by shear-flexure interaction. Fig. \ref{fig:distributions} presents the distribution of key design parameters for each failure mode.  Further details on the database can be obtained elsewhere \cite{tubitak1001report}.

\subsection{Definition of backbone curves}
\begin{figure}[!h]
\centering
\includegraphics[scale=0.4]{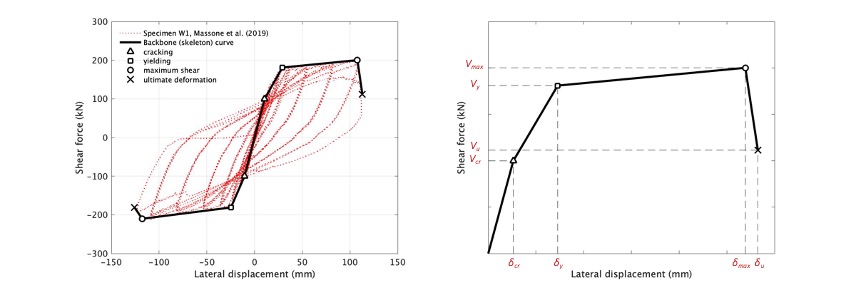}
\caption{(a) Load – deformation response \cite{massone2019experimental}; (b) backbone curve definition}
\label{fig:loaddef}
\end{figure}

 Backbone curve of each specimen is obtained using cyclic base shear-top lateral displacement loops by taking the average of responses in positive and negative regions, as shown in Fig. \ref{fig:loaddef}(a) for a representative specimen. The deformation components are not separated; that is, the total deformation included contributions from flexure, shear, and slip of the reinforcement. 
 
Backbone curves include shear force and displacement at four critical stages, namely: cracking, yielding, reaching shear strength, and reaching ultimate displacement. The shear ($V$\textsubscript{cr}) and displacement ($\delta$\textsubscript{cr}) at the cracking point are related through the elastic wall properties; therefore, the following seven backbone variables are investigated in this study: shear at cracking point ($V$\textsubscript{cr}), shear and displacement at yielding point ($V$\textsubscript{y} and $\delta$\textsubscript{y}), shear and displacement at shear capacity ($V$\textsubscript{max} and $\delta$\textsubscript{max}), as well as shear and displacement at displacement capacity ($V$\textsubscript{u} and $\delta$\textsubscript{u}).

\begin{figure}[!h]
\includegraphics[width=\textwidth]{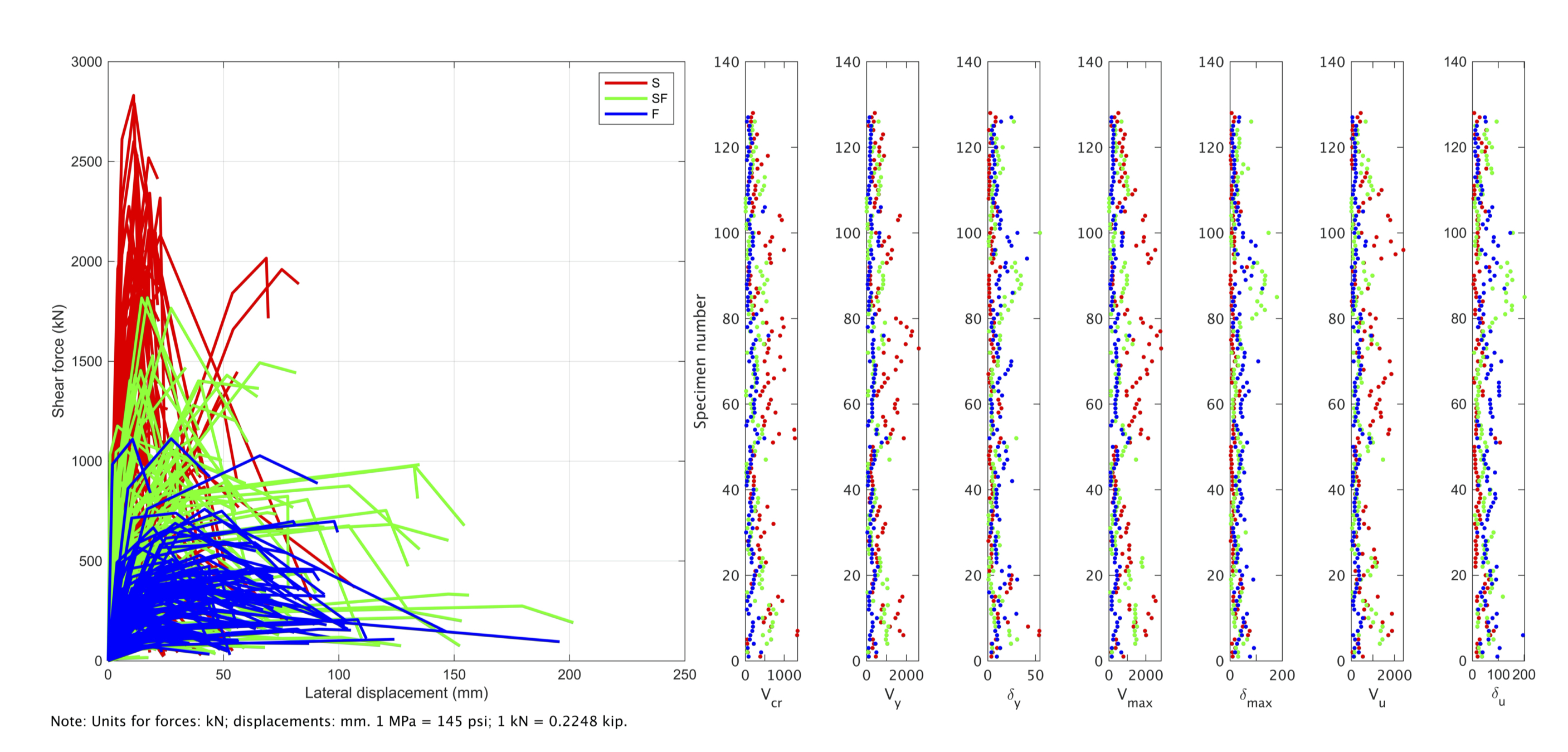}
\caption{(a) Backbone curves of 384 specimens in the database, (b) Distribution of seven backbone curve variables}
\label{fig:5}
\end{figure}

 Shear and flexure behaviors can be assessed together in Fig. \ref{fig:5}(a) based on backbone curves plotted with identifying colors for different failure modes. Observations indicate that deformation capacity is influenced by the level of shear strength as also shown in literature \cite{deger2019empirical,oesterle1981reinforcement,Priestley1994,peeratc72}. Wall shear strength degrades with increasing ductility, leading to shear failure (red backbone curves), whereas, higher displacements can be reached under low shear levels (blue backbone curves).

Fig. \ref{fig:5}(b) demonstrates the distribution of the seven backbone variables in the wall database. The seven backbone variables are compared to ASCE 41-17 recommended modeling parameters (a, b, c, d, e, f, g values shown in Fig.\ref{fig:ascebc}) in Tables  \ref{table:1} and \ref{table:2} for shear-controlled and flexure-controlled walls, respectively. Note that dominant wall types in Table \ref{table:1} and Table \ref{table:2} are identified based on reported failure modes of the specimens, as data for walls controlled by flexure based on ASCE 41-17 classification criteria (i.e., shear-controlled if $h$\textsubscript{w}/$l$\textsubscript{w}$<$1.5, flexure-controlled if $h$\textsubscript{w}/$l$\textsubscript{w}$>$3) is too scarce to make an assessment. Comparisons indicate that ASCE 41 modeling parameters achieve a fairly good agreement for walls subjected to low axial load with the following exceptions for the sake of conservatism: (i) ASCE 41 neglects the significant increase in the slope after yield point (between $V$\textsubscript{y} and $V$\textsubscript{max}), (ii) ASCE 41 residual strength recommendations ($V$\textsubscript{u}) are more much lower for shear-controlled walls, (iii) ASCE 41 does not address the data dispersion, therefore, can be too conservative in some cases. For other wall types, higher discrepancies are observed except for well-predicted $f$,  (i.e. $V$\textsubscript{cr}/$V$\textsubscript{y}) value for shear-controlled walls. It is noted that ASCE 41 does not provide modeling parameters for walls that have shear-flexure interaction.

\begin{table}[h!]
\caption{Comparisons of database backbone values with ASCE 41-17 recommendations for shear-controlled walls}
\centering
\begin{tabular}{l l l l}
\hline
Parameter & 
$\frac{(A_s-A_s')f_yE+P}{t_wl_wf_c}$ & ASCE 41-17 & Database\\
\hline
$V$\textsubscript{max}/$V$\textsubscript{y}& 
$\le$ $0.5$ & $1.0$      & $1.43$$\pm$$0.37$\\
 & \textgreater0.5& 1.0      & 1.18$\pm$0.05\\ 
$V$\textsubscript{u}/$V$\textsubscript{max} (c)& 
$\le$ 0.5& 0.2      & 0.65$\pm$0.29\\
& \textgreater0.5& 0.0      & 0.75$\pm$0.13\\ 
$V$\textsubscript{cr}/$V$\textsubscript{y} (f)& 
$\le$ 0.5& 0.6 & 0.58$\pm$0.17\\
 & \textgreater0.5& 0.6 & 0.61$\pm$0.23\\ 
$\delta_{max}/h$ (d)& $\le$ 0.5& 1.0      & 0.97$\pm$0.48\\
& \textgreater0.5& 0.75    & 1.21$\pm$0.19\\ 
$\delta_{u}/h$ (e) & $\le$ 0.5& 2.0     & 1.84$\pm$1.19\\  & \textgreater 0.5& 1.0      & 2.05$\pm$0.46\\ 
$\delta_{y}/h$ (g)& $\le$ 0.5& 0.4      & 0.42$\pm$0.3\\ & \textgreater 0.5&   0.4      & 0.57$\pm$0.08\\ 
 \hline
\end{tabular}
\label{table:1}
\end{table}

\begin{table}[h!]
\caption{Comparisons of database backbone values with ASCE 41-17 recommendations for flexure-controlled walls}
\small
\scriptsize
\centering
\begin{tabular}{l l l l l l l l}
\hline
$\frac{(A_s-A_s')f_yE+P}{t_wl_wf_c}$  & $\frac{V}{t_wl_wf'_cE}$  & \begin{tabular}[l]{@{}l@{}}Confined \\ boundary\end{tabular} & ASCE 41-17             & Database      & ASCE 41-17 & Database    \\
\hline
&   &   & \multicolumn{2}{l}{$V_{max}$/$V_y$} & \multicolumn{2}{l}{c = $V_u$/$V_{max}$}\\
\textless{}=0.1     & \textless{}=4    & Yes     & 1.0  & 1.29 & 0.75 & 0.75  \\
\textless{}=0.1     & \textgreater{}=6 & Yes     & 1.0  & 1.38  &  0.4 & 0.70   \\
\textgreater{}=0.25 & \textless{}=4    & Yes     & 1.0  & 1.37  &   0.6  & 0.75   \\
\textgreater{}=0.25 & \textgreater{}=6 & Yes & 1.0  & 1.22  &   0.3 & 0.73     \\
\textless{}=0.1     & \textless{}=4    & No & 1.0 & 1.27  &   0.6 & 0.90           \\
\textless{}=0.1     & \textgreater{}=6 & No      & 1.0   & No data & 0.3 & No data  \\
\textgreater{}=0.25 & \textless{}=4    & No & 1.0 & 1.34          &   0.25     & 0.62    \\
\textgreater{}=0.25 & \textgreater{}=6 & No   & 1.0  & No data & 0.2 & No data \\
\hline
$\frac{(A_s-A_s')f_yE+P}{t_wl_wf_c}$  & $\frac{V}{t_wl_wf'_cE}$  & \begin{tabular}[l]{@{}l@{}}Confined \\ boundary\end{tabular} & ASCE 41-17             & Database      & ASCE 41-17 & Database    \\
\hline
&   &    &
\multicolumn{2}{l}{a = $\theta_{max}$-$\theta_y$} & \multicolumn{2}{l}{b = $\theta_{u}$ -$\theta_{y}$} \\
\textless{}=0.1     & \textless{}=4    & Yes     &   0.015      & 0.011         &   0.020     & 0.019          \\
\textless{}=0.1     & \textgreater{}=6 & Yes     &   0.010      & 0.011         &   0.015     & 0.019          \\
\textgreater{}=0.25 & \textless{}=4    & Yes     &   0.009      & 0.008         &   0.012     & 0.016          \\
\textgreater{}=0.25 & \textgreater{}=6 & Yes &   0.005     & 0.012         &   0.010     & 0.025          \\
\textless{}=0.1     & \textless{}=4    & No  &   0.008      & 0.013         &   0.015    & 0.020          \\
\textless{}=0.1     & \textgreater{}=6 & No  & 0.006   & No data & 0.010 & No data  \\
\textgreater{}=0.25 & \textless{}=4    & No & 0.003 & 0.007   & 0.005     & 0.019     \\
\textgreater{}=0.25 & \textgreater{}=6 & No &  0.002 & No data & 0.004 & No data\\ \hline     
\end{tabular}
\label{table:2}
\end{table}

\section{Background on Gaussian Process Regression}
Gaussian Process Regression (GPR) is obtained by generalizing linear regression. The GPR is  a collection of random variables, with a joint Gaussian distribution, hence is a probabilistic multivariate regression method. Let us  $D=\{(x_i,f_i), i=1,\ldots,n\}$ refer to a given training set, where $f_i = f(x_i)$ is the value of the function at the sample $x_i$. The Gaussian Process (GP), a generalization of the Gaussian probability distribution, is represented as follows: 
\begin{equation}
f(\mathbf{x})\; \mathtt{\sim} \; GP(m(\mathbf{x_i}),k(\mathbf{x}_i,\mathbf{x}_j))
\end{equation}
where $\mathbf{x}$ and $k(\mathbf{x}_i,\mathbf{x}_j)$ refer to mean and covariance function, respectively. The covariance matrix express the similarity between the pair of random variables and determined by means of  squared exponential (SE) kernel function given as 
\begin{equation}
k(\mathbf{x}_i,\mathbf{x}_j)={exp}(||\mathbf{x}_i-\mathbf{x}_j||^2/(2\sigma^2)) 
\end{equation}
where $\sigma$ is the kernel width that needs to be optimized.  The predictive function,  $f_*i$ can be obtained for a given test sample $X*$, as following, 
\begin{eqnarray}
  p(\mathbf{f}_* | \mathbf{X}_*, \mathbf{X}, \mathbf{f})  =  N(\mathbf{f}_* | \mathbf{\mu}_*, \mathbf{\sigma}_*)
\end{eqnarray} 
where
\begin{eqnarray}
  \mu_*  &=& \underline{k}_*^T. [K(X,X)+\sigma_n^2 I]^{-1}\underline{y}  \\ 
  \sigma_*^2 &=& k(\underline{X}_*,\underline{X}_*)- \underline{k}_*^T [K(X,X) + \sigma_n^2]^{-1} \underline{k}_*.
\end{eqnarray}
$\underline{k}_*$ is a vector holding the  covariance values between training samples $X$ and the test samples  $\underline{X}_*$. The  $\sigma_n^2$ and $\sigma_x^2$ refer to the noise variance and the confidence measure associated with the model to output, respectively. 

\section{Implementation of Gaussian Process Regression and Results}

\subsection{Experimental Setup}
The key wall properties (Fig.\ref{fig:distributions}) and the critical backbone variables are designated as input variables (features) and output variables for the regression analyses, respectively.  Note that the problem is a multi-output regression since multiple outputs are needed to be estimated.Training and test datasets are randomly selected as disjoint sets from the database at a ratio of 90\% and 10\%, respectively, and this random splitting is repeated a hundred times. The purpose of a hundred trials is to assess the proposed method's generalization capability and robustness. Because of the randomization, training/test pairings vary in each realization, resulting in a hundred different learning models.  The GPR-based models are constructed on each training dataset,  whereas  the  accuracy  of  the  resulting  models  is evaluated  based on associated test dataset. The validation results are then averaged over a hundred trials for each output. It should be noted that experiments with fewer training datasets (e.g. 80\% for training and 20\% for testing) are also considered; however, a large number of trials are overtrained, implying that the amount of the training set is not sufficient to generate a reliable learning model. Based on this experience, the size of the training set is increased to 90\% of all data.

The kernel parameters of the GPR method are automatically tuned using gradient-based optimization routines to maximize the marginal likelihood. 

The performance of models is evaluated based on three criteria: the mean coefficient of determination (R${}^{2}$), relative root mean square error (RelRMSE), and the prediction accuracy, that is, the ratio of predicted to actual values (P/A). The experiments are conducted based on a hundred trials to ensure the reliability and robustness of the proposed model, and the averaged results are reported.  All the codes are developed in Matlab \cite{MATLAB} on a personal computer with Dual Core Intel Core i7 CPU (3.50GHz), 16GB RAM, and MacOS operating system.

Data pre-processing has been shown as a useful technique to improve the quality of machine learning algorithms by transforming the data to a more informative format \cite{jiang2008can}. Among several data transformation techniques available, Box-cox and log transformation methods are applied to the outputs. The original form of Box-Cox transformation equation \cite{box1964analysis} is as shown in Eq. \ref{eq:boxcox}.

\begin{equation}
\label{eq:boxcox}
y(\lambda )=\left\{\begin{array}{c} {\begin{array}{cc} {\frac{y^{\lambda } -1}{\lambda } ,} & {{\rm if}\lambda \ne 0} \end{array}} \\ {\begin{array}{cc} {\log (y),} & {{\rm if}\lambda =0} \end{array}} \end{array}\right. 
\end{equation}

where $y$ is the data value and $\lambda$ is a value between -5 and 5. The optimal $\lambda$ is the one that transforms the data set to the best approximation of normal distribution. The log transformation corresponds to $\lambda$ equals zero. It is observed that the use of log transformation does not improve the results, whereas Box-Cox (optimal $\lambda$ = -0.3) leads to an increase in model accuracy for all outputs, particularly for 
$\delta$\textsubscript{y}, $\delta$\textsubscript{max}, and $\delta$\textsubscript{u} 
(Table \ref{table:boxcox}). 

\begin{table}[h!]
\caption{Effect of data transformation on GPR model results (Averaged values over 100 trials).}
\centering
\begin{tabular}{ccccc} 
 \hline  
{Output} &  \multicolumn{2}{c}{No Transformation}  & \multicolumn{2}{c}{Box-Cox Transformation} \\ 
  \textbf{ } &  R${}^{2}$ &  RelRMSE & R${}^{2}$ & RelRMSE\\ [0.5ex] 
 \hline
$V_y$ &  0.80 &  0.36  & 0.84 & 0.33 \\ 
  $\delta_y$ &  0.72 & 0.55 &  0.82 & 0.50 \\
  $V_{max}$ &  0.87 & 0.28 &  0.89 & 0.28  \\
  $\delta_{max}$ &  0.84 & 0.43 &  0.91 &  0.33  \\
    $V_u$ &  0.57 & 0.64&  0.62 &  0.59  \\ 
  $\delta_u$ & 0.77 & 0.40 &  0.89 &  0.29   \\ [0.5ex] 
 \hline
\end{tabular}
\label{table:boxcox}
\end{table}

\subsection{Experimental Results and Discussion}
\begin{figure}[!h]
\centering
\includegraphics[scale=0.35]{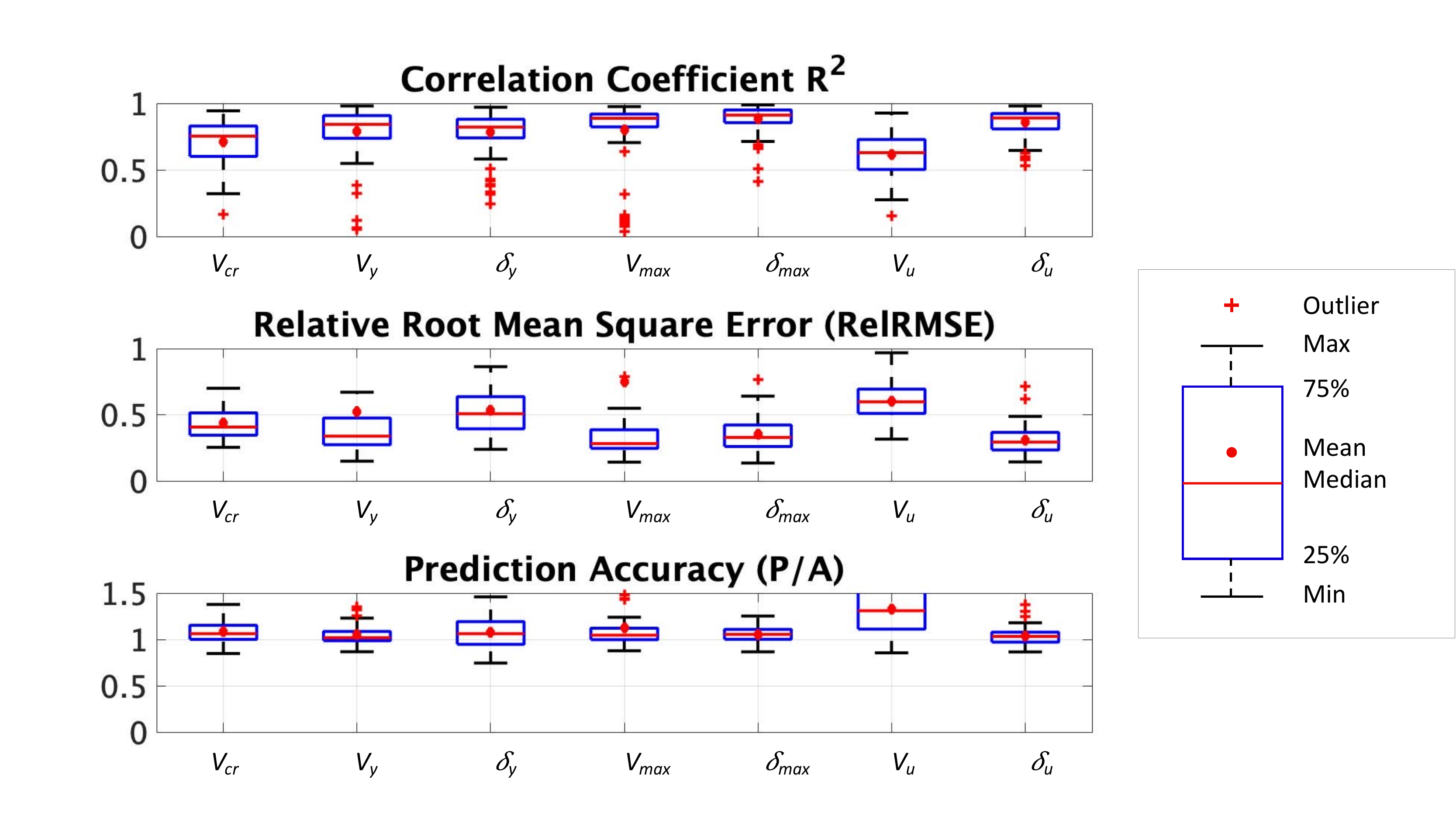}
\caption{GPR model performances for the backbone variables over 100 trials.}
\label{fig:boxplots}
\end{figure}

\textcolor{blue}{}{Prediction accuracy and correlation values based on machine learning experiments are summarized in Table \ref{table:performance100} for each output. The results reveal that the GPR method achieves better correlations (R${}^{2}$\textgreater{0.8}) for walls that failed in shear (S) or shear-flexure interaction (SF), and the relatively low correlation coefficients (mean R${}^{2}$ around 0.65) reduce the performance of models.} It is noted that modeling parameters in the horizontal axis of backbone curves ($\delta$\textsubscript{y}, $\delta$\textsubscript{max}, $\delta$\textsubscript{u}) were also studied in terms of chord rotation and curvature for flexure-controlled walls and drift ratio for the  shear-controlled walls; however, correlation coefficients were much lower. Therefore, the results presented in Fig. \ref{fig:boxplots} focus on walls that reached their shear strength, that is, those controlled by shear or shear-flexure interaction, and the prediction models are proposed for such walls for the seven backbone variables.

\begin{table}[h!]
\small
\caption{Averaged performances of the GPR-based models over 100 trials }
\centering
\begin{tabular}{cccccc} 
 \hline 
Output & & \begin{tabular}[c]{@{}c@{}}R${}^{2}$\\ (median)\end{tabular} &  \begin{tabular}[c]{@{}c@{}}Mean \newline\\ RelRMSE\end{tabular} & \begin{tabular}[c]{@{}c@{}}Mean\\ P/A\end{tabular} &  \begin{tabular}[c]{@{}c@{}}Std. Dev.\\ P/A\end{tabular} \\
 \hline
\multirow{2}{2em}{\textcolor{blue}{$V_{cr}$}} & \textcolor{blue}{S, SF}     & \textcolor{blue}{0.71}    &   \textcolor{blue}{0.44}    &  \textcolor{blue}{1.08}     &   \textcolor{blue}{0.47}\\
& \textcolor{blue}{All}    & \textcolor{blue}{0.67}    &   \textcolor{blue}{0.54}    &  \textcolor{blue}{1.07}     &   \textcolor{blue}{0.46} \\
\multirow{2}{2em}{$V_{y}$}	& S, SF     &	0.84	&	0.33	&	1.02	&	0.29	\\
                       & All    & 0.79    &   0.52    &  1.06     &   0.41 \\
\multirow{2}{2em}{$\delta_y$} & S, SF 	&	0.82  &	0.5	&	1.06	&	0.46	\\
                       & All    & 0.64    &   0.55    &  1.11     &   0.66 \\
\multirow{2}{2em}{$V_{max}$}	& S, SF 	&	0.89	&	0.28	&	1.04	&	0.26	\\
                        & All       & 0.80    &   0.69    &  1.09     &   0.57 \\
\multirow{2}{2em}{$\delta_{max}$}	& S, SF 	&	0.91	&	0.33	&	1.05	&	0.38	\\
                            & All       & 0.79    &   0.42    &  1.03     &   0.57 \\
\multirow{2}{2em}{$V_u$}	& S, SF 	&	0.62	&	0.59	&	1.3	&	0.99	\\
                        & All       & 0.79    &   0.42    &  1.03     &   0.57 \\
\multirow{2}{2em}{$\delta_u$}	& S, SF 	&	0.89	&	0.29	&	1.03	&	0.29	\\                              & All       & 0.79    &   0.33    &  1.05     &   0.36 \\ [0.5ex]
 \hline
\end{tabular}
\label{table:performance100}
\end{table}

Due to overtraining, trials with poor generalization capability such as those with the highest correlation in training but very low correlation in testing, are omitted, resulting in the mean values shown with the red diamond in Fig. \ref{fig:boxplots} and the median values tabulated in Table \ref{table:performance100}.The results indicate that coefficient of determination values are over 80$\%$ with low relative RMSE values (around 0.3) for all seven backbone variables; except for the shear force corresponding to ultimate deformation (i.e.$V$\textsubscript{u}), the mean R${}^{2}$ is around 60$\%$ with higher error. As seen in the boxplots, prediction accuracy is very close to 1.0 with  low standard deviations for all seven backbone variables, except for the shear force at ultimate deformation. The lower performance for $V$\textsubscript{u} implies that input-output relation can not be established for this output, potentially because some specimens reached their ultimate deformation, whereas, for the others, ultimate deformation is estimated at 80$\%$ of shear strength, as mentioned earlier. This may have changed the structure of the problem, resulting in relatively poor predictions.

\begin{figure}[!h]
\centering
\includegraphics[scale=0.4]{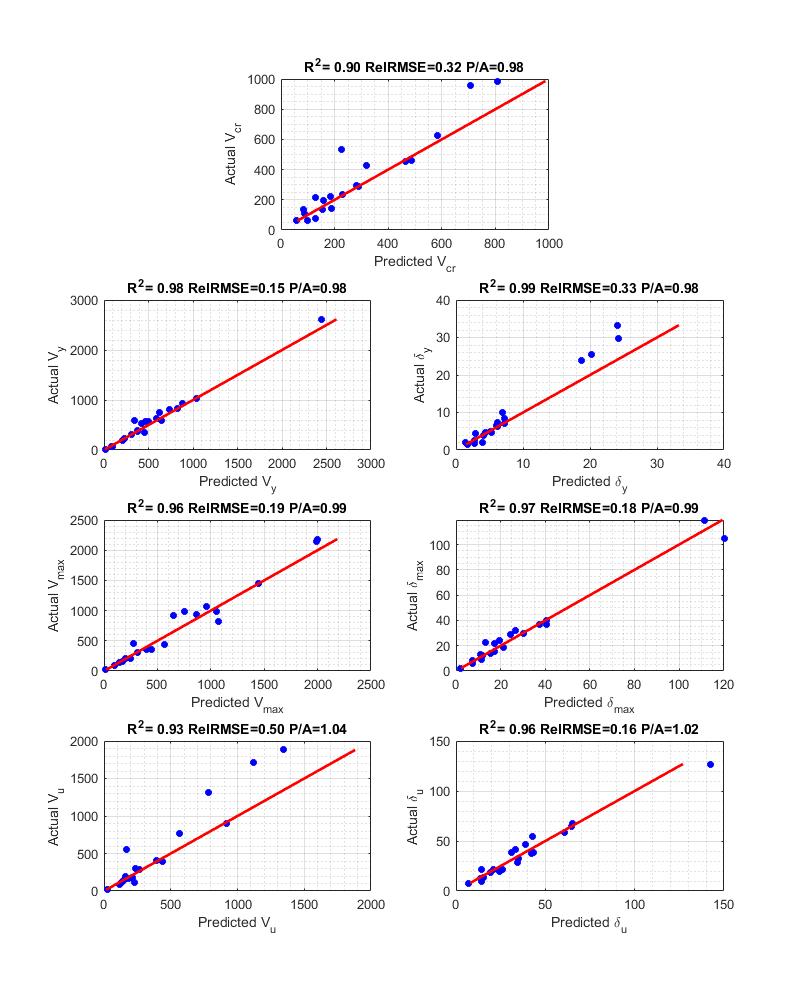}
\caption{Scatter plots for correlations of the proposed model for each backbone variable}
\label{fig:scatters}
\end{figure}

The correlation between the actual and predicted values is assessed based on the proposed GPR-based predictive model that achieves the highest R${}^{2}$, the P/A closest to 1.0, and the lowest error simultaneously (Fig. \ref{fig:scatters}) for each backbone variable. The scatter plots indicate that data are closely distributed around the line of y = x, implying that the proposed models are able to make accurate predictions when compared to the actual values. Results demonstrate that the proposed GPR-based models achieve excellent performance in predicting shear strength $V$\textsubscript{max} and corresponding displacement ($\delta$\textsubscript{max}) with highest coefficients of determination (R${}^{2}$ equals to 0.96 and 0.97, respectively) and smallest relative RMSE. The proposed model for ultimate deformation ($V$\textsubscript{u}) also achieves good agreement with the actual data even though mean values over 100 trials were lower as this output is more data-dependent. 
  
To evaluate the influence of each wall property (i.e. feature, Fig.\ref{fig:distributions}) on the model output, optimized lengthscale hyperparameters of the kernel function are used in the GPR model as each hyperparameter controls the spread of each input. The hyperparameter optimization is carried out by computing the partial derivatives of the negative log-marginal likelihood function with respect to the given parameter. As small values show greater relevance, their inverses are used for feature ranking. The most influential features are axial load ratio (${P}/{({{A}_{g}}} f_{c}^{'})$) and aspect ratio (${h_w}/{l_w}$) for force-related backbone variables ($V$\textsubscript{cr}, $V$\textsubscript{y}, $V$\textsubscript{max}, and $V$\textsubscript{u}), whereas  shear span ratio (${M}/{Vl}_{w}$) is more critical for the displacement variables ($\delta$\textsubscript{y}, $\delta$\textsubscript{max}, $\delta$\textsubscript{u}).
  
The performance of the proposed models is also compared to ASCE 41 modeling recommendations based on 20 unseen test data in Table \ref{table:asceGPRcomp}. Mean and standard deviation of the predicted/actual (P/A) values are assessed as performance indicators for each backbone variable, where ASCE 41 values are back-calculated from the normalized modeling parameters. Mean P/A values are close to 1.0 for all seven backbone variables for the GPR-based models, whereas higher errors are observed based on ASCE 41. The standard deviation of the P/A is also much smaller for the GPR-based models. Results indicate that the proposed models outperform the traditional ASCE 41 approach for a wider spectrum of walls (versus only low axial loads, as discussed in Table 1 with higher accuracy.  

% Please add the following required packages to your document preamble:
% \usepackage{booktabs}
% \usepackage{multirow}
\begin{table}[h!]
\caption{Performance of the proposed model for each backbone variable}
\centering
\begin{tabular}{@{}ccccc@{}}
\hline
\multicolumn{1}{c}{Output} & \multicolumn{2}{c}{ASCE 41} & \multicolumn{2}{c}{GPR}   \\
\multicolumn{1}{c}{} & P/A mean  & P/A dispersion  & P/A mean & P/A dispersion \\ 
\hline

$V_y$  & 2.17  &   2.41  &  0.97 &   0.16 \\
$\delta_y$ & 1.41 & 1.08 &  0.98 & 0.32 \\
$V_{max}$  & 1.28 & 0.52 & 0.99 & 0.18 \\
$\delta_{max}$ & 0.90 & 0.44 & 0.99 & 0.17 \\
$V_u$ & 0.40 & 0.28 & 1.04 & 0.39 \\
$\delta_u$  & 1.27 & 0.55 & 1.02 & 0.18    \\
\hline
\end{tabular}
\label{table:asceGPRcomp}
\end{table}

Machine learning experiments revealed that unusually thin ($l$\textsubscript{w}/$t$\textsubscript{w}$>$20) shear walls and slender walls ($h$\textsubscript{w}/$l$\textsubscript{w}$>$3.0) decrease the correlation and the prediction accuracy. The former is expected as typical $l$\textsubscript{w}/$t$\textsubscript{w} for shear walls is reported as 13.3 \cite{wallace2012behavior} and reaches $l$\textsubscript{w}/$t$\textsubscript{w}=19 for around 350cm-long shear walls \cite{magna2014non} which corresponds to the maximum wall length in the database. The latter factor decreasing correlations ($h$\textsubscript{w}/$l$\textsubscript{w}) can be associated with the typical wall behavior, that is, squat wall responses are governed by shear and slender wall responses are dominated by flexural behavior. However, experimental results showed that squat walls may fail in flexure if adequate reinforcement and detailing are provided \cite{salonikios1999cyclic}, whereas slender walls may fail in shear (web crushing) \cite{sittipunt2001cyclic}. Therefore, backbone relations of slender walls need to be analyzed separately even though they happened to fail in shear. 

As noted above, the proposed GPR-based backbone curve models are generalized and can be used regardless of failure modes. However, results demonstrate that the predictions are more reliable especially for shear-controlled and shear-flexure controlled walls. It is worth advising the potential users of the proposed GPR-based models in that identification of dominant wall behavior (failure mode) based on ASCE 41 may lead to conservative assessments. For example, some moderate aspect ratio walls (1.5$<$ $h$\textsubscript{w}/$l$\textsubscript{w}$<$3) would be classified as shear-flexure controlled even though they would actually fail in flexure, creating a false impression that backbones of such walls can be predicted based on the proposed GPR-based models. Therefore, prior to the use of the proposed models for a new shear wall specimen, a more accurate failure mode prediction \cite{mangalathu2020data,degertaskin} is recommended. 

It is also noted that other machine-learning methods including k-nearest neighbour regression (KNN) \cite{Cover:2006:NNP:2263261.2267456}, least absolute shrinkage and selection operator (LASSO) \cite{tibshirani96regression}, neighbourhood component analysis (NCA) \cite{goldberger2004neighbourhood}, random forest \cite{DBLP:journals/ml/Breiman01}, regularized logistic regression (RLR) \cite{hosmer2013applied}, and support vector regression (SVR) \cite{Vapnik1998} were also utilized in this study in an attempt to obtain higher correlations and accuracies; however, none of them were as accurate as the GPR method.

\section{Summary and Conclusions}
Backbone curves are critical component modeling tools in element-based nonlinear modeling to represent the response under reversal drift cycles. It  is essential to simulate the hysteretic response as accurately as possible to achieve realistic assessments as nonlinear static and dynamic analyses are conducted based on such curves.

This study aims to address this need by developing predictive models for backbone curves based on a comprehensive shear wall experimental database using a powerful machine learning method, that is Gaussian Process Regression (GPR). Seven backbone curve variables associated with four limit states, namely: cracking, yielding, peak shear strength, and ultimate deformation capacity are predicted in terms of twelve wall design properties (e.g. concrete compressive strength, reinforcement details, wall geometry). To achieve this, the backbone curve variables and wall design properties are designated as outputs and inputs, respectively. The database is divided into train and test data sets, whereas the performances of the proposed models are evaluated based on statistical metrics (e.g. coefficient of determination, relative root mean square error, prediction accuracy) based on test data sets. The prediction performances of the proposed GPR models are also compared to traditional modeling approaches (ASCE 41-17) and shown to be superior.

For best performance, the proposed models are recommended for walls that satisfy the following conditions: (i) wall behavior is controlled by shear or shear-flexure interaction, (ii) $l$\textsubscript{w}/$t$\textsubscript{w}$\le$20, and (iii) $h$\textsubscript{w}/$l$\textsubscript{w}$\le$3. 

The results reveal that one can predict the shear force - displacement relations, thus, the existing capacity and seismic performance of RC shear walls, close to accurate using the proposed GPR-based models. This study will contribute to the literature by providing tools that achieve dual objectives of high accuracy performance and computational efficiency to ensure more realistic performance evaluations.

\section{Acknowledgements}
	The project has been supported by funds from the Scientific and Technological Research Council of Turkey (TUBITAK) under Project No: 218M535. Opinions, findings, and conclusions in this paper are those of the authors and do not necessarily represent those of the funding agency. 

\newpage
\section*{Notation}
\begin{table}[!h]
\scriptsize
\begin{tabular}{l l} 
${{A}_{g}}$  &=\space\space\space\space  gross cross section area of wall  \\
${\delta}_{cr}$ &=\space\space\space\space  cracking displacement \\ 
${\delta}_{y}$ &=\space\space\space\space  yield displacement  \\ 
${\delta}_{max}$ &=\space\space\space\space displacement when peak shear is reached\\ 
${\delta}_{u}$ &=\space\space\space\space  ultimate displacement  \\ 
$\varepsilon$ &=\space\space\space\space  the noise term \\
$E\left( X|Y \right)$ &=\space\space\space\space expected value of random variable X given Y \\
$f_{c}^{'}$ &=\space\space\space\space  concrete compressive strength  \\
${f}_{ybl}$ &=\space\space\space\space  longitudinal boundary reinforcement yield strength  \\
${f}_{yl}$ &=\space\space\space\space  longitudinal web reinforcement yield strength  \\
${f}_{ysh}$ &=\space\space\space\space  transverse boundary reinforcement yield strength  \\
${f}_{yt}$ &=\space\space\space\space  transverse web reinforcement yield strength  \\
${h}_{w}$ &=\space\space\space\space  wall height \\ 
${h_w/l_w}$  &=\space\space\space\space  aspect ratio \\
$\mathbf{I}$  &=\space\space\space\space  the identity matrix\\
${l}_{w}$ &=\space\space\space\space  wall length \\
$\mathcal{L}$ &=\space\space\space\space  the loss function \\
$m$  &=\space\space\space\space  number of specimens\\
${M}/{Vl}_{w}$  &=\space\space\space\space  shear-span ratio \\
$n$  &=\space\space\space\space  number of design parameters\\
${P}/{({{A}_{g}}} f_{c}^{'})$  &=\space\space\space\space  axial load ratio  \\
$\mathbf{P}\left( A|B \right)$ &=\space\space\space\space probability of event A given event B occured \\
${s}$  &=\space\space\space\space  stirrups spacing  \\
${t}_{w}$ &=\space\space\space\space  wall thickness  \\
${{\rho }_{bl}}$ &=\space\space\space\space  longitudinal boundary reinforcement ratio  \\
${{\rho }_{l}}$ &=\space\space\space\space  longitudinal web reinforcement ratio  \\
${{\rho }_{sh}}$ &=\space\space\space\space  transverse boundary reinforcement ratio\\
${{\rho }_{t}}$ &=\space\space\space\space  transverse web reinforcement ratio  \\
$\mathbf{w}$ &=\space\space\space\space  weighing vector of features  \\
$\mathbf{X}$ &=\space\space\space\space  $m\times n$ dimensional matrix holding all the features \\
${{x}_{i}}$ &=\space\space\space\space   $i^{th}$ feature value of the ML model\\
$\mathbf{y}$ &=\space\space\space\space  $m$ dimensional output vector \\
${V}_{cr}$ &=\space\space\space\space  shear force at cracking \\ 
${V}_{y}$ &=\space\space\space\space  shear force at yielding \\ 
${V}_{max}$ &=\space\space\space\space  peak shear force \\ 
${V}_{u}$ &=\space\space\space\space  shear force when ultimate displacement is reached  \\ 
\end{tabular}
\end{table}

\newpage
\clearpage

%%%%%%%%%%%%%%%%%%%%%%%
\bibliographystyle{unsrt}
\bibliography{BackboneML}

\end{document}